\documentclass[showkeys,showpacs,twocolumn]{revtex4-1}
\usepackage{dcolumn}
\usepackage{amsmath}
\usepackage{amssymb}
\usepackage{graphicx}
\usepackage{bm}

\newcommand\be{\begin{equation}}
\newcommand\ee{\end{equation}}
\newcommand{\bea}{\begin{eqnarray}}
\newcommand{\eea}{\end{eqnarray}}
\newcommand\non{\nonumber}
\newcommand\eps{\epsilon}
\newcommand{\tr}{\mathop{\rm Tr}}
\newcommand\half{\frac{1}{2}}

\def\NN{\mathop{N}}

\def \outlineby #1#2#3{\vbox{\hrule\hbox{\vrule\kern #1%
\vbox{\kern #2 #3\kern #2}\kern #1\vrule}\hrule}}%

\newtheorem{theorem}{Theorem}[section]
\newtheorem{lemma}{Lemma}[section]

\def\sgn{{\rm \sgn}}

\begin{document}

\bibliographystyle{ieeetr}


\title{The invalidity of a strong capacity for a quantum channel with memory}
\author{Tony Dorlas}
\email{dorlas@stp.dias.ie}
\affiliation{
School of Theoretical Physics\\
Dublin Institute for Advanced Studies \\ 10 Burlington Road, Dublin 4, Ireland}
\author{Ciara Morgan}
\email{cqtciara@nus.edu.sg}
\affiliation{Centre for Quantum Technologies\\ National University of Singapore \\ 3 Science Drive 2, Singapore 117543}

\begin{abstract}

The strong capacity of a particular channel can be interpreted as
a sharp limit on the amount of information which can be
transmitted reliably over that channel. To evaluate the strong
capacity of a particular channel one must prove both the direct
part of the channel coding theorem and the strong converse for the
channel. Here we consider the strong converse theorem for the
periodic quantum channel and show some rather surprising results.
We first show that the strong converse does not hold in general
for this channel and therefore the channel does not have a strong
capacity. Instead, we find that there is a scale of capacities
corresponding to error probabilities between integer multiples of
the inverse of the periodicity of the channel. A similar
scale also exists for the random channel.

\end{abstract}

\keywords{channel coding theorem, classical capacity, quantum channels with memory}

\pacs{03.37.Hk, 89.70.Kn}

\maketitle

\section{Introduction}

The full channel coding theorem provides a limit on the rate at
which a sender can communicate an encoded message to a receiver,
such that the probability of a decoding error at the receiver's
side decays exponentially in the number of channel uses. The
theorem is comprised of two parts: the direct part of the theorem,
which refers to the construction of the code, and the converse to
the theorem. The direct part of the quantum channel coding theorem
states that using $n$ copies of the channel, we can code with an
exponentially small probability of error at a rate $R =\frac{1}{n}
\log |\mathcal{M}|$, provided $R < C$ in the asymptotic limit,
where $\cal M$ denotes the set of possible codewords to be
transmitted over the channel and $C$ denotes the capacity of the
channel.
If the rate at which classical information is transmitted over a
quantum channel exceeds the capacity of the channel, i.e. if $R>C$,
then the probability of decoding the information correctly goes to
zero in the number of channel uses. The latter is known as the strong
converse to the channel coding theorem. The \emph{weak} converse,
on the other hand, states that if $R>C$, then the probability of
decoding the information correctly is bounded away from $1$, i.e.
the error probability does not tend to zero, whatever encoding/decoding
scheme is used.\\
Shannon \cite{Shannon48} first proposed the theorem for classical
discrete memoryless channels and the first rigorous proof of the
direct part of the theorem was provided by Feinstein \cite{Feinstein}
and the strong converse by Wolfowitz \cite{Wolfowitz}. \\
However, it was observed that the existence of the strong converse,
and therefore strong capacity, for other types of classical channels
does not always hold \cite{WolfowitzNoCap}. See Ahlswede \cite{Ahlswede06}
for a more complete discussion of converse results for various types of classical channels. \\
The strong converse to the channel coding theorem for memoryless \emph{classical-quantum} channels with product state inputs was determined
independently by Winter \cite{Winter99} and by Ogawa and Nagaoka \cite{ON99}.
Their result implies that every memoryless discrete classical-quantum channel
has a strong capacity which provides a sharp upper-bound on the rate at which
classical information can be transmitted over this type of channel using
product states.\\
Recent results include a proof by Bjelakovi\'c and Boche
\cite{BB08, BB09} of a full coding theorem for the discrete
memoryless compound classical-quantum channel. Wehner and K\"onig
\cite{KW09} proved the fully general strong converse theorem for a
family of channels, that is, they proved that the strong converse
theorem
holds for a family of quantum channels even in the case when entangled state inputs are allowed.\\
In this article we relax the assumption that the communication channel
in question is memoryless and we concentrate on a particular quantum
channel with memory, that is, a channel with correlations between
successive channel uses. In our case the correlations between successive
uses of the channel can be described by a Markov chain. Communication
channels with memory are widely considered to be more realistic than
memoryless channels since real-world channels may not exhibit independence
between successive errors and correlations are common. Noise correlations
are also necessary for certain models of quantum communication \cite{Bose03}.
See for example Kretschmann and Werner \cite {KW05} and Mancini \cite{Mancini06}
for models of quantum memory channels.\\
The article is organised as follows. We introduce notation, necessary definitions
and define the quantum periodic channel in Section \ref{prelim}.
In Section \ref{NoConverse} we prove that the periodic channel does not
have a strong capacity. The observation relies on a result which is proved
in Appendix \ref{app}, involving a particular instance of a periodic channel
and consequently the strong converse does not hold in general for the periodic channel.
In Section \ref{RandomCSect} we remark on a scale of capacities for the random channel.
We then state and prove the main result involving a scale of capacities for the channel.\\
Note that $\log$ is understood to be taken to the base $2$
throughout the article.

\section{Preliminaries}\label{prelim}

We begin by introducing some notation. A memoryless channel is given by a
completely positive trace-preserving (CPT) map
${\Phi}: {\mathcal{B}}({\cal H}) \to {\cal B}({\cal K})$, where
${\cal B}({\cal H})$ and ${\cal B}({\cal K})$ denote the states on
the input and output Hilbert spaces ${\cal H}$ and $\cal K$, respectively.\\
Equivalently, we can describe a \emph{classical-quantum} channel,
here also denoted $\Phi$, as a mapping from the classical message
to the output state of the channel on $\mathcal{B}(\mathcal{K})$
as follows, \be \Phi : \mathcal{X} \mapsto
\mathcal{B}(\mathcal{K}), \ee where the message is first encoded
into a sequence belonging the set
$\mathcal{X}^n$, where $\cal X$ represents the input alphabet.\\
We can combine the two mapping descriptions as follows. We wish to
send classical information in the form of quantum states over a
quantum channel $\Phi$.
A (discrete) memoryless quantum channel, $\Phi$, carrying classical
information can be thought of as a map from a (finite) set, or alphabet,
$\cal X$ into $\cal B(\cal K)$, taking each $x \in \cal X$ to
$\Phi_{x}= \Phi(\rho_x)$, where the input state to the channel is
given by $\{ \rho_x\}_{x \in \cal{X}}$ and each $\rho_x \in \cal B (\cal H)$.
Let $d=\dim(\cal H)$ and $a=|\cal X|$.\\
For a probability distribution $P$ on the input alphabet
$\mathcal{X}$, the average output state of a channel $\Phi$ is
given by \be P \sigma = \sum_{x \in \mathcal{X}} P(x)
\Phi(\rho_x). \ee The conditional von Neumann entropy of $\Phi$
given $P$ is defined by \be S(\Phi|P) = \sum_{x \in \mathcal{X}}
P(x) S(\Phi(\rho_x)), \hfill \ee and the mutual information
between the probability distribution $P$ and the channel $\Phi$ is
defined as follows, \be\label{mutInf} I(P;\Phi)= S(P \sigma) -
S(\Phi|P). \ee

An \emph{$n$-block code} for a quantum channel $\Phi$ is a pair
$(C^n,E^n)$, where $C^n$ is a mapping from a finite set of
messages $\cal M$, of length $n$, into $\mathcal{X}^n$, i.e. a
sequence $x^n \in \cal X$ is assigned to each of the $|\cal M |$
messages, and $E^n$ is a POVM, i.e. a quantum measurement, on the
output space $\mathcal{K}^{\otimes n}$ of the channel
$\Phi_{x^n}^{n}$. The maximum error probability of the code
$(C^n,E^n)$ is defined as \be p_e(C^n,E^n) = {\rm max}\, \{ 1 -
\tr (\Phi_{C^n(m)}^{n} E^{n}_m) : m \in \cal M \}. \ee The code
$(C^n,E^n)$ is called an $(n, \lambda)$-code, if $p_e(C^n,E^n)
\leq \lambda$. The maximum size $| \cal M |$ of an $(n,
\lambda)$-code is denoted $N(n,\lambda)$. Define an finite
alphabet $\cal X$ and sequences $x^n = x_1 , \dots, x_n \in
\mathcal{X}^n$ and let \be N(x \big|x^n) = \big| \{ i \in \{1,
\dots, n\} : x_i = x \} \big| \ee for $x \in \mathcal{X}$. The
\emph{type} of the sequence $x^n$ is given by the empirical
distribution $P_{x^n}$ on $\cal X$ such that \be\label{distP}
P_{x^n}(x) = \frac{N(x \big|x^n)}{n}. \ee Clearly, the number of
types is upper bounded by $(n+1)^a$, where $a= \big| \cal X
\big|$.

\subsection{Coding theorem and strong converse}

The \emph{strong} capacity of a particular channel provides a
\emph{sharp} threshold on the rate at which information may be
transmitted over that channel with \emph{exponentially decreasing}
probability of decoding error in the number of channel uses.
In order to establish a strong capacity for a particular channel
one must prove both existence of a capacity achieving code and the
strong converse.\\
The direct part of the coding theorem for memoryless quantum channels
with product-state inputs was determined independently by Holevo \cite{Hol98} and Schumacher and
Westmoreland \cite{SW97}. Winter \cite{Winter99} and
Ogawa and Nagaoka \cite{ON99} independently proved the strong
converse for memoryless quantum channels.\\
In Section \ref{NoConverse} we require a version of the strong
converse theorem proved by Winter \cite{Winter99} which holds for
a single codeword type. We therefore provide this version (Lemma
\ref{SameTypeTheorem}) below, following both the direct part and
strong converse theorems for memoryless classical quantum channels
as stated and proved in \cite{Winter99}.

\begin{theorem}{\rm (Direct part)}\\
For all $\lambda \in (0,1)$ and $\delta > 0$ there exists
$n_0(\lambda,\delta) \in \NN$ such that for all $n \geq n_0$ and
every classical quantum channel $\Phi$ and probability
distribution $P$ on $\cal{X}$, there exists an $(n,\lambda)$-code
such that the number of messages satisfies \be | \mathcal{M}_n |
\geq 2^{n (\chi^*(\Phi) - \delta)},
\ee where the Holevo capacity $\chi^*$ is given by \be
\chi^*(\Phi) = \sup_P I(P;\Phi) \ee the supremum being over all
probability distributions $P$ on $\cal X$.
\end{theorem}

\begin{theorem}{\rm (Strong converse)}\\
For all $\lambda \in (0,1)$ and all $\delta > 0$ there exists
$n_1(\lambda,\delta)$ such that for all $n \geq n_1$ and every
memoryless classical quantum channel $\Phi$ and the number of
messages of an $(n,\lambda)$-code is bounded by \be
|\mathcal{M}_n| \leq 2^{n (\chi^*(\Phi) + \delta)}.
\ee
\end{theorem}

\textbf{Remark.} \textit{Winter in fact proved a stronger version
of these theorems in which $\delta$ is replaced by a constant
times $1/\sqrt{n}$}.

In the following we follow the approach of Winter
(\cite{Winter99}, Theorem 13) in which the strong
converse is derived from a bound on the number of codewords of a given
\emph{type} $P$:
\begin{lemma}\label{SameTypeTheorem}{\rm (Single-type strong converse)}\\
For $\lambda \in (0,1)$ and $\delta > 0$ there exists
$n_1(\lambda,\delta)$  such that for $n \geq n_1$, every
$(n,\lambda)$-code for which all codewords are of the same type
$P$, \be |\mathcal{M}_{n,P}| \leq  2^{n (I(P; \, \Phi) + \delta)}.
\ee
\end{lemma}

The strong converse follows immediately from this lemma using the
fact that the number of types is upper bounded by $(1 + n)^a$ (see \cite{CK11} Lemma 2.2).
\\

\textbf{Remark.} \textit{In contrast to the strong converse where
the decoding error goes to $1$ exponentially in the number of
channel applications if $R>C$, the \emph{weak} converse states
that if $R>C$, then the probability of decoding the information
correctly is bounded away from $1$.}

\subsection{Quantum channels with classical memory}\label{Markov}

Next, we provide definitions needed to describe quantum channels
with classical memory \cite{Norris}. Let $I$ denote a countable
set and let $\lambda_i = \mathbb{P}(X=i)$, where $X$ is a random
variable taking values in the state space $I$. Let $Q$ denote a
transition matrix, with entries labeled $q_{j|i}$. A discrete time
random process denoted $X_n$ can be considered to be a Markov
chain with transition matrix $Q$ and initial distribution
$\lambda$, if and only if the following holds for $i_0, \dots,
i_{n-1} \in I$,
\bea  &\mathbb{P}&(X_0 =i_0, X_1 =
i_1, \dots , X_{n-1} = i_{n-1}) \non \\ &=& \lambda_{i_{0}}
q_{i_{1}|i_{0}} q_{i_{2}|i_{1}} \cdots q_{i_{n-1}|i_{n-2}}.
\eea
In \cite{DD09Markov} Datta and Dorlas analyse a quantum
channel of length $n$ with Markovian noise correlations, first
defined by Bowen and Mancini \cite{BM04}, as follows \be
\label{GenMarkov} \Phi^{n}(\rho^{n}) = \hspace{-3mm} \sum_{i_0
\dots i_{n-1}} \hspace{-3mm} q_{i_{n-1}|i_{n-2}} \dots q_{i_1|i_0}
 \lambda_{i_{0}} (\Phi_{i_0} \otimes \cdots \otimes
\Phi_{i_{n-1}}) (\rho^{n}) \ee where $q_{j|i}$ are the elements
of the transition matrix of a discrete-time Markov chain, and $\{
\lambda_i\}$ represents an invariant distribution on the Markov chain.\\
In Section \ref{NoConverse} we analyse a particular channel with
classical memory, namely the periodic channel. We describe this channel below.\\
A periodic channel acting on an $n$-fold input state can be
described as follows \be\label{perC} \Phi^{n} \left( \rho^{n}
\right) = \frac{1}{L} \sum_{i=0}^{L-1} \left( \Phi_i \otimes
\Phi_{i+1} \otimes \cdots\otimes \Phi_{i+n-1} \right) \left( \rho
^{n} \right), \ee where $\Phi_i$ are CPT maps acting on the same
Hilbert space and the index is cyclic, modulo the period $L$, i.e.
$\Phi_{i+L} = \Phi_i$.  In this case the elements of the
corresponding transition matrix are given by $q_{j|i} =
\theta_{i,j}$, where \be \theta_{i,j}=
\begin{cases} 1, & \text{if $j=i+1 \mod L$}
\\
0, &\text{otherwise.}
\end{cases}
\ee The product-state capacity of the channel, denoted $C_{p}$ is
given by \be\label{cap_PerSect} C_{p} \left( \Phi \right) =
\frac{1}{L} \sup_{P} \sum_{i=0}^{L-1} I(P; \Phi_i). \ee The proof
of direct part of the channel coding theorem for the periodic
quantum channel is provided in Appendix B of \cite{CMThesis}. This
is in fact a special case of the main result proved by Datta and
Dorlas in \cite{DD09Markov}. Note that the proof of the direct
part of the coding theorem for this channel makes use of a
\emph{preamble} to the code which the receiver uses upon receipt to
determine which branch of the channel was selected.

Another channel of the general type (\ref{GenMarkov}) is the
random channel. It is given by \be\label{randomC} \Phi^{n} \left(
\rho^{n} \right) = \sum_{i=1}^{M} q_i \,\Phi_i^{\otimes n}\left(
\rho ^{n} \right), \ee where $\Phi_i$ ($i=1,\dots,M$) are CPT maps
acting on the same Hilbert space and $q_1,\dots,q_M$ is a
probability distribution.  In this case the elements of the
corresponding transition matrix are given by $q_{j|i} =
\delta_{ij}$. It was shown in \cite{DD07CC} that the product state
capacity of this channel is given by \be\label{cap_random} C_{p}
\left( \Phi \right) = \sup_{P} \min_{i=1}^M I(P; \Phi_i). \ee We
will remark on this channel, which like the periodic channel has
long-term memory, in Section~\ref{RandomCSect}.

\section{Channel without a strong converse}\label{NoConverse}

The strong converse for the periodic quantum channel does not hold
in general because the following inequality holds
\be\label{capIneq} C_{p} < \overline{C_{p}}, \ee where,
\be\label{CBar} \overline{C_{p}} = \frac{1}{L} \sum_{i=0}^{L-1}
\sup_{P} I(P, \Phi_i). \ee The strict inequality above can be
shown explicitly for a periodic channel consisting of two branches
of qubit amplitude-damping channels (see Appendix \ref{app} below
for detailed proof). On the other hand, equality for expression
(\ref{capIneq}) can be shown to hold for a periodic channel with
depolarising channel branches \cite{DM09}.

Let us now investigate whether we can prove a full coding
theorem for rates $R$ such that \be\label{compRate} C_{p} < R < \overline{C_{p}}.
\ee We first define the average probability of error as follows
\be\label{AvErrorProb} \overline{p_e} = \frac{1}{L} \sum_{i=0}^{L-1}
p_{e}^i \leq \lambda, \ee
where $p_e^i$ denotes the probability of error for the $i$-th channel branch.

Our coding strategy is as follows. We choose a code i.e. a
particular encoding and decoding scheme, suitable for a particular
channel branch labeled by the index $i \in \{ 0, \dots, L-1\}$.
Here a \lq branch' is defined as one term in the sum (\ref{perC})
i.e. $\Phi_i^{(n)} = \Phi_i \otimes \Phi_{i+1} \otimes \dots
\otimes \Phi_{i+n-1}$. According to the coding theorem for memoryless
channels, there is a code with error probability tending to zero
for this branch with rate $R$. Indeed, for each $j$ there exists a
probability distribution $P_j$ of states optimising $\chi_j^* =
\sup_P I(P;\Phi_j)$ and we can choose states from a typical
subspace for these distributions, which can be interlaced at the
positions $j-i+kL$, where $k \in [\frac{n}{L} -1]$. The probability of choosing a particular
branch correctly is given by $\frac{1}{L}$ and therefore the
probability of error approaches \be p_e = \frac{L-1}{L} < 1. \ee
We thus have a $\lambda$-code for all $\lambda > 1-\frac{1}{L}$.
In particular, the error probability is bounded away from 1, and the \emph{strong} converse does not hold.

On the other hand the strong converse does hold for $R >
\overline{C_p}$. Indeed, the codewords can be decomposed into
sub-codewords corresponding to the different stages of a period:
$x^n = (x_0^n,\dots, x_{L-1}^n)$, where the components of the
$x_i^n$ are understood to be interlaced in $x^n$. We distinguish
types $P_0, \dots, P_{L-1}$ for the sub-codewords. Then we have an
analogue of the single-type strong converse given by Lemma~\ref{SameTypeTheorem}:
\begin{lemma} \label{MultipleTypes}
For $\lambda \in (0,1)$ and $\delta > 0$ there exists
$n_1(\lambda,\delta)$  such that for $n \geq n_1$, every
$(n,\lambda)$-code for which all sub-codewords are of the same
type $P_0,\dots,P_{L-1}$, given that the $i$-th branch is
selected, \be |\mathcal{M}_{n,P_0,\dots,P_{L-1}}| \leq
2^{\frac{n}{L} \sum_{k=0}^{L-1} (I(P_k; \, \Phi_{i+k}) + \delta)}.
\ee
\end{lemma}
Clearly, for the complete channel, it follows that the number of
codewords such that the sub-codewords are of types
$P_0,\dots,P_{L-1}$, satisfies \be |{\cal
M}_{n,P_0,\dots,P_{L-1}}| \leq 2^{\frac{n}{L} \sum_{k=0}^{L-1} (\sup_{P}
I(P;\Phi_{i+k})+\delta)}. \ee Summing over the types, we obtain
the strong converse.

We can conclude that the strong converse holds for rates $R > \overline{C_p}$.

\section{A scale of capacities}

The above obviously raises the question if smaller error
probabilities can be attained for smaller rates, but still above
$C_p$. For this, we define a \lq pair capacity' $C_p^{\,(2)}$ as
follows: \be\label{pairCap} C_p^{\,(2)} = \frac{1}{2L} \max_{0 \leq i_1 < i_2 < L}
\sum_{k=0}^{L-1} \hspace{-1mm}\sup_P \left( I(P;\Phi_{i_1+k}) +
I(P;\Phi_{i_2+k}) \right). \ee Suppose the maximum is attained at
a certain pair $(i_1,i_2)$. With probability $2/L$, one of the two
branches $i_1$ or $i_2$ is chosen. We attach a preamble to the
code as in the proof of the product-state capacity of the periodic channel (\ref{cap_PerSect}). If, for example, the
branch $i_1$ is selected by the channel, the receiver can
determine that this is the case by measuring the preamble, and can
then choose states for each value of $k$ from the typical space
corresponding to the maximising distribution $P_k$ for the CPT map
$\Phi_{i_1+k}$. This constitutes an encoding with rate given by
the average of the mutual informations $I(P;\Phi_{i_1+k})$ for
$k=0,\dots,L-1$, which is greater than or equal to the pair capacity $C^{\,(2)}$ given by Equation (\ref{pairCap}).
We have thus constructed a $\lambda$-code for $\lambda > 1-\frac{2}{L}$.

On the other hand, let $R > C^{\,(2)}_p$ and suppose that $(C^n,E^n)$ is a
sequence of  $(n,\lambda)$-codes with $\lambda < 1-\frac{1}{L}$,
and assume that
\be \label{high_rate} \frac{1}{n} \log |{\cal
M}_n| \geq R > C_p^{\,(2)}. \ee

First note that we may assume that
the number of codewords with sub-codewords of types $P_0, \dots
P_{L-1}$ is bounded by
\be \frac{1}{n} \log |{\cal
M}_{n,P_0,\dots,P_{L-1}}| \leq \frac{1}{L} \sum_{k=0}^{L-1}
(I(P_k, \Phi_{i+k})+\delta) \ee
for some \emph{fixed} $i=0,\dots,L-1$. Indeed,
otherwise, by Lemma \ref{MultipleTypes}, $p^i_e > \lambda$ for all $i$ and
hence $p_e > \lambda$.

We now claim that for every other $j \neq
i$, and $\eps > 0$ small enough, \be \label{rateclaim} \frac{1}{n}
\log |{\cal M}_{n,P_0,\dots,P_{L-1}}| > \frac{1}{L}
\sum_{k=0}^{L-1} (I(P_k, \Phi_{j+k})+\eps). \ee

If this were not the case then the pair capacity for a single type $P_k$ can be written as
\bea \frac{1}{n} \log |{\cal M}_{n,P_0,\dots,P_{L-1}}|
&\leq& \frac{1}{2L} \sum_{k=0}^{L-1} ( I(P_k, \Phi_{i+k}) \non \\
&+& I(P_k, \Phi_{j+k}) +\delta ) \eea

and hence \bea \frac{1}{n} \log |{\cal
M}_{n,P_0,\dots,P_{L-1}}| &\leq& \frac{1}{2L} \sum_{k=0}^{L-1} (
\sup_P  \{I(P, \Phi_{i+k}) \non \\ &+& I(P,\Phi_{j+k})\}
+\delta ). \eea
Summing over the types $P_0,\dots,
P_{L-1}$ leads to a contradiction with (\ref{high_rate}).

Now, expression (\ref{rateclaim}) implies with Lemma~\ref{MultipleTypes}
that, if the $j$-th branch is selected by the channel, then the
error probability $p^j_e > 1-\eta$ for any $\eta > 0$. Since with
probability $1-\frac{1}{L}$ one of the branches $j$ other than $i$
is selected, we conclude that the error probability $p_e >
\left(1-\frac{1}{L}\right) (1-\eta) > \lambda$ if $\eta <
1-\frac{1}{L} - \lambda$ is small enough.

It is now clear that this argument can be generalised to prove:
\begin{theorem} Define, for $r=1, \dots,L$ a \textbf{scale} of capacities $C_p^{(r)}$ by
\be C_p^{(r)} = \frac{1}{rL} \max_{0 \leq i_1 < \dots < i_r < L}
\sum_{k=0}^{L-1} \sup_P \sum_{m=1}^r I(P;\Phi_{i_m+k}). \ee (Note
that $C_p^{(1)} = C_p$ and $C_p^{(L)} = \overline{C_p}$.) Then, if
$\lambda > 1-\frac{r}{L}$ and $R < C_p^{(r)}$, there exists a
sequence of $(n,\lambda)$-codes with rate $R$. Conversely, if
$\lambda < 1-\frac{r -1}{L}$, there exists no sequence of
$(n,\lambda)$-codes with rate $R > C_p^{(r)}$. \end{theorem}

\section{The random channel} \label{RandomCSect}

The situation for the random channel is similar, but more
complicated due to the fact that different branches can have
different probabilities $q_i$. We can in general distinguish break
points at values of the error probability given by \be q(\Delta) =
\sum_{i \in \Delta} q_i, \qquad \Delta \subset \{1,\dots,M\}. \ee
We have an analogue of the detailed theorem for periodic channels
above:
\begin{theorem} Define, for $\Delta \subset \{1,\dots,M\}$ a \textbf{scale}
of capacities $C_p^{\Delta}$ by \be C_p^{\Delta} = \sup_P \min_{i
\in \Delta} I(P;\Phi_{i}). \ee Then, if $\lambda > 1-q(\Delta)$
and $R < C_p^{\Delta}$, then there exists a sequence of
$(n,\lambda)$ codes with rate $R$.

For the converse to the theorem, we introduce another scale as follows:
\be
\overline{C_p^\Delta} = \sup_P \max_{i \in \Delta} I(P;\Phi_i).
\ee
Then, if $\lambda < 1-q(\Delta)$ there exist no $(n,\lambda)$-codes with
rate $R > \overline{C_p^\Delta}$.
\end{theorem}

The situation is less clear-cut than it seems, however. In fact, not every $q(\Delta)$
is necessarily a point of discontinuity for the capacity, because
$C_p^\Delta$ is in general not monotonic in the probabilities
$q(\Delta)$!

\section{Discussion}\label{discussion}

One of the most surprising and interesting results which has
emerged from Shannon Theory is the observation that the
\emph{strong} information-carrying capacity of a memoryless
channel is independent of the upper bound on the maximum error
probability of that channel,
usually denoted $\lambda$. The independence of the parameter
$\lambda$ is crucial to the existence of a so-called \emph{strong}
capacity for the channel \cite{Wolfowitz}.\\
The dependency of some channel capacities on this parameter
$\lambda$, including non-stationary discrete memoryless classical
channels, led to the definition of a capacity function
\cite{Ahlswede06}. Note that recently Ahlswede \cite{Ahlswede10}
proved that the
capacity functions can now be thought of as so-called capacity-sequences.\\
For the case of the quantum periodic channel, and also the random
channel, we have shown that an analogous parameter-dependent
capacity can be defined, which takes the form of a scale of
capacities applicable for various ranges of the error parameter.

\textbf{Note.} It appears that similar results to ours were
obtained by Datta, Hsieh and Brand\~{a}o \cite{DHB11}, using different methods.

\appendix
\section{The periodic channel with amplitude-damping channel branches}\label{app}

The qubit amplitude-damping channel acting on the state $\rho =
\left(
\begin{array}{cc} a & b \\ \overline{b} & 1-a \end{array} \right)$ is given by
\be \Phi_{amp}(\rho) = \left(\begin{array}{cc}
a + (1-a) \gamma & b\sqrt{1 - \gamma}  \\
\overline{b}\sqrt{1- \gamma} & (1-a)(1- \gamma)   \\
\end{array} \right).
\ee The expression for the product-state capacity of the qubit
amplitude-damping channel is given as follows,
\begin{eqnarray}\label{amp1}
&\chi& \hspace{-1mm} \left(\Phi_{amp}(\{p_j,\rho_j\})\right) \non
\\ &=& S \left[ \sum_j \left(
\begin{array}{cc}
p_j\left(a_j+(1-a_j)\gamma \right) & p_j b_j\sqrt{(1 - \gamma)}  \non \\
p_j {\overline b}_j \sqrt{(1 - \gamma)} & p_j(1-a_j)(1 - \gamma)   \\
\end{array} \right) \right]\\
&-& \sum_j p_j\, S \left(
\begin{array}{cc}
a_j+(1-a_j)\gamma & b_j\sqrt{1 - \gamma}  \\
{\overline b}_j \sqrt{1 - \gamma} & (1-a_j)(1 - \gamma)   \\
\end{array} \right).
\end{eqnarray}
We now investigate whether the following equation holds for a
periodic channel with two amplitude-damping channel branches
\be\label{conj_amp} \half \sup_{\{p_j, \rho_j\}} \sum_{i=0}^{1}
\chi_i(\{p_j, \rho_j\})  = \half \sum_{i=0}^{1}   \sup_{\{p_j,
\rho_j\}} \chi_i(\{p_j, \rho_j\}). \ee Let $\gamma_0$ and
$\gamma_1$ represent the error parameters for two
amplitude-damping channels $\Phi_0$ and $\Phi_1$ respectively. We
have argued \cite{DM08} that the Holevo quantity for the qubit
amplitude-damping channel can be increased using an ensemble
containing two mirror image pure states each with probability
$\half$. Using this minimal ensemble we investigate both sides of
Equation (\ref{conj_amp}), for a periodic channel with two qubit
amplitude-damping channel
branches.\\
Clearly the \emph{left hand side} of Equation (\ref{conj_amp})
will be attained for a single parameter which we denote by
$a_{max}$. However, the \emph{right hand side} of Equation
(\ref{conj_amp}) cannot be obtained by a single $a_{max}$.
Instead, the supremum for each channel will be attained at a
different value of the input state parameter $a$. We denote by
$a_{max_0}$ and $a_{max_1}$ the state parameter that achieves the
product-state capacity for the channels $\Phi_0$ and $\Phi_1$
respectively. Let $\chi_0(a)$ and $\chi_1(a)$ denote the Holevo
quantities of the channels $\Phi_0$ and $\Phi_1$, respectively.
Denoting $x_i=\sqrt{1-4\gamma_{i}\left(1 - \gamma_{i} \right)(1-
a^2) }$ the eigenvalues for each of the amplitude-damping channels
can be written as \be \lambda_{amp_i\pm} = \frac{1}{2} \left(1 \pm
\sqrt{1-4\gamma_i(1-\gamma_i)(1-a)^2} \right). \ee the values for
$a_{max_0}$ and $a_{max_1}$ can be determined by separately
solving the following equation for each channel
\begin{eqnarray} \label{detRHSa0} \frac{d \chi_i (a)}{d a } &=&
( 1- \gamma_i) \ln \left(
\frac{(1-a)(1-\gamma_i)}{a + (1-a)\gamma_i} \right) \non \\
&+& \frac{2 \gamma_i (1-\gamma_i)(1-a)}{x_i} \ln\left(
\frac{1+x_i}{1-x_i}
\right) \non \\
&=& 0. \end{eqnarray} Let
$\chi^*_{avg}(\gamma_0,\gamma_1,a_{max_0},a_{max_1})$ denote the
average of the supremum of the Holevo capacities of the channels
$\Phi_0$ and $\Phi_1$, i.e., \be
\chi^*_{avg}(\gamma_0,\gamma_1,a_{max_0},a_{max_1}) = \half
\left(\chi_0^*(a_{max_0}) + \chi_1^*(a_{max_1}) \right). \ee It is
not difficult to show that \be \chi^*(\gamma_0=1,\gamma_1,a_{max})
= \chi^*_{avg}(\gamma_0=1,\gamma_1,a_{max_0},a_{max_1}). \ee
Similarly, we can show that
$$\chi^*(\gamma_0,\gamma_1=1,a_{max}) =
\chi^*_{avg}(\gamma_0,\gamma_1=1,a_{max_0},a_{max_1}).$$ Next, we
show separately for a) $\gamma_i = 0$ and for b) $0 < \gamma_i <
1$ that the following inequality holds \be
\chi^*(\gamma_0,\gamma_1,a_{max}) <
\chi^*_{avg}(\gamma_0,\gamma_1,a_{max_0},a_{max_1}). \ee

\begin{enumerate}
\item[a)]\label{a} Taking $\gamma_0 = 0$, the expression
$\chi^*(\gamma_0,\gamma_1,a_{max})$ becomes \begin{eqnarray}
\chi^*(\gamma_0=0,\gamma_1,a_{max}) &=&  H_{bin}(a_{max1}) \non \\
&+&  H_{bin}((1-a_{max1})(1 - \gamma_1)) \non \\
&-&   S\left( \Phi_1\left( \rho_{amax} \right) \right).
\end{eqnarray} Denoting $\chi^*_{avg}(\gamma_0,
\gamma_1,a_{max_0},a_{max_1})$ by $\chi^*_{avg}\left( \gamma_1
\right)$ the right hand side becomes \begin{eqnarray}
 \chi^*_{avg}\left( \gamma_1, a_{max1} \right) &=&  H_{bin}(a_{max1}) \non \\
&+&  H_{bin}((1-a_{max1})(1 - \gamma_1))  \non \\
&-&   S\left( \Phi_1\left( \rho_{amax_1} \right) \right).
\end{eqnarray} Clearly, $a_{max_0} = \half$. To show that $a_{max} <
a_{max_1}$, we must show that $ \frac{d }{da}\sum_i \chi_i (a) <0$
at $a=a_{max_1} = \half$.

For $\gamma_0 = 0$ the Holevo quantity of the channel $\Phi_0$
becomes \be \chi_0 (a)= S  \left(\begin{array}{cc}
a  & 0  \\
0 & (1-a)   \\
\end{array} \right) - S (\rho).
\ee But $\rho$ is a pure state and therefore $S(\rho) = 0$.
Therefore, from Equation (\ref{detRHSa0}), \be \frac{d \chi_0
(a)}{d a} = \ln \left( \frac{(1- a)}{a} \right). \ee We have
previously shown that the maximising state parameter for the
amplitude-damping channel is achieved at $a \geq \half$
\cite{DM08}. We are considering the case where $\gamma_0 \neq
\gamma_1$, i.e. $\gamma_1 \neq 0$, therefore $a_{max_1} > \half$.
The expression $\chi_0(a)$ now represents the binary entropy,
$H(a)$, and is therefore maximised at $a= \half$. It was shown
above that the entropy $S(a)$ is a strictly concave function for
$\gamma_0 = 0$ and $\chi_0 (a)$ is therefore decreasing at $a =
a_{max_1}$.

The capacity $\chi_1^*(a)$ is achieved at $a = a_{max_1}$.
Therefore $\frac{d \chi_1 (a)}{da}$ is equal to zero at this
point.

We can  now conclude that $\frac{d }{da}\sum_i \chi_i (a) <0$ when
$a=a_{max_1}$ and therefore \begin{eqnarray}
\chi^*(\gamma_0=0,\gamma_1,a_{max}) <
\chi^*_{avg}(\gamma_0=0,\gamma_1,a_{max_0},a_{max_1}). \non \\
\end{eqnarray}

\item[b)]\label{b} We now show that an inequality exists between
the expressions $\chi^*(\gamma_0,\gamma_1,a_{max})$  and
$\chi^*_{avg}(\gamma_0,\gamma_1,a_{max_0},a_{max_1})$ for fixed
$\gamma_0$, such that $0 < \gamma_0 <1$.

In \cite{DM08} we proved that if $\gamma_0 < \gamma_1$, then
$\chi(\gamma_0) > \chi(\gamma_1)$ and therefore $a_{max_0} <
a_{max_1}$. Therefore, $\frac{d \chi_0 (a)}{da} <0$ at
$a=a_{max_1}$ and $a_{max} < a_{max_1}$. Similarly, if $\gamma_0 >
\gamma_1$, then $a_{max_0} > a_{max_1}$ and $\frac{d
 \chi_0 (a)}{da} >0$ at $a=a_{max_1}$ and $a_{max} > a_{max_1}$.

As a result, $a_{max}$ will always lie in between $a_{max_0}$ and
$a_{max_1}$. We have previously shown \cite{DM08} that the Holevo
quantity for the qubit amplitude-damping channel is concave in its
single state parameter. Therefore $a_{max} > \tilde{a}$, where
$\tilde{a}$ is the parameter value associated with
$\chi^*_{avg}(\gamma,\gamma_1,a_{max_0},a_{max_1})$, i.e. $\sum_i
\sup_a \chi_i(a) = \chi^*_{\gamma_0,\gamma_1} (\tilde{a})$.
This proves that $\chi^*(\gamma_0,\gamma_1,a_{max}) <
\chi^*_{avg}(\gamma,\gamma_1,a_{max_0},a_{max_1})$.
\end{enumerate}

In conclusion, if $\gamma_0 = 1$ or $\gamma_1=1$, then $a_{max} =
a_{max_0}$ or $a_{max} = a_{max_1}$ respectively and
$\chi^*(\gamma_0,\gamma_1,a_{max}) =
\chi^*_{avg}(\gamma,\gamma_1,a_{max_1},a_{max_1})$. However, if
$\gamma_0, \gamma_1 \neq 1$, then
$\chi^*(\gamma_0,\gamma_1,a_{max}) <
\chi^*_{avg}(\gamma,\gamma_1,a_{max_1},a_{max_1})$.

\bibliography{Bib}

\end{document}